\documentclass[twocolumn]{article}
\usepackage{inputenc}
\usepackage{amsmath,amssymb}
\usepackage{fullpage,hyperref}
\usepackage{verbatim}
\usepackage{graphicx}
\def\bt{{\bf t}}
\title{Function-Valued Traits in Evolution} 
\author{Pantelis Z. Hadjipantelis \footnote{Centre for Complexity Science and Department of Statistics, University of Warwick, Coventry CV4 7AL, UK}, Nick S. Jones\footnote{Department of Mathematics, Imperial College London, SW7 2AZ}, John Moriarty\footnote{School of Mathematics, University of Manchester, Oxford Road, Manchester M13 9PL, UK} , David A. Springate\footnote{Faculty of Life Sciences, University of Manchester, Oxford Road, Manchester M13 9PT, UK} , Christopher G. Knight\footnotemark[\value{footnote}]}
\date{}

\begin{document}
\maketitle

\begin{abstract}
Many biological characteristics of evolutionary interest are not scalar variables but continuous functions. Given a dataset of function-valued traits generated by evolution, 
we develop a practical statistical approach to infer ancestral function-valued traits, and estimate the generative evolutionary process. We do this by combining dimension reduction and phylogenetic Gaussian process regression, a nonparametric procedure which explicitly accounts for known phylogenetic relationships. We test the methods' performance on simulated function-valued data generated from a stochastic evolutionary model. The methods are applied assuming that only the phylogeny and the function-valued traits of taxa at its tips are known. Our method is robust and applicable to a wide range of function-valued data, and also offers a phylogenetically aware method for estimating the autocorrelation of function-valued traits.
\end{abstract}


Abbreviations:

ICA: Independent Components Analysis

IPCA: Independent Principal Components Analysis

MLE: Maximum Likelihood Estimator

OU: Ornstein-Uhlenbeck

PCA: Principal Components Analysis

PGP: Phylogenetic Gaussian Process

PGPR: Phylogenetic Gaussian Process Regression

\section{Introduction}
\label{sec:Intro}

The number, reliability and coverage of evolutionary trees are growing rapidly \cite{Maddison07,TreeFam}. However, knowing organisms' evolutionary relationships through phylogenetics is only one step in understanding the evolution of their characteristics \cite{Yang12}. Three issues are particularly challenging. The first is limited information: empirical information is typically only available for extant taxa, represented by tips of a phylogenetic tree, whereas evolutionary questions frequently concern unobserved ancestors deeper in the tree. The second is dependence: the available information for different organisms in a phylogeny is not independent since a phylogeny describes a complex pattern of non-independence; observed variation is a mixture of this inherited variation and specific variation \cite{Cheverud85}. The third is high dimensionality: the emerging literature on function-valued traits \cite{Kirkp89,TFPG12,Stinchcombe12} recognises that many characteristics of living organisms are best represented as a continuous function rather than a single factor or a small number of correlated factors. Such characteristics include growth or mortality curves \cite{Pletcher99}, reaction-norms   \cite{Kingsolver01} and distributions \cite{Zhang11}, where the increasing ease of genome sequencing has greatly expanded the range of species in which distributions of gene \cite{Moss11} or predicted protein \cite{Knight04} properties are available. Therefore, a function-valued trait is defined as a phenotypic trait that can be represented by a continuous mathematical function \cite{Kingsolver01}.

Previous work \cite{JM12} proposed an evolutionary model for function-valued data  $d$ related by a phylogeny $\bf{T}$.  The data are regarded as observations of a phylogenetic Gaussian Process (PGP) at the tips of  $\bf{T}$. That work shows that a PGP can be expressed as a stochastic linear operator $X$ on a fixed set $\phi$ of basis functions (independent components of variation), so that
\begin{equation} d = X^T \phi
\label{lip}
\end{equation}
However, the study does not address the linear inverse problem of obtaining estimates $\hat{\phi}$ and $\hat{X}$ of $\phi$ and $X$: our first contribution in this paper is to provide an approach to this problem in section \ref{sec:DimensionRed} via independent principal components analysis (IPCA \cite{Yao12}). 

We refer to $X$ as the {\em mixing matrix}, and to the $(i,j)$th entry of $X$ as the {\em mixing coefficient} of the $i$th basis function at the $j$th taxon. It is these mixing coefficients that we model as evolving. For each fixed value of $i$, the $X_{ij}$ are correlated (due to phylogeny) as $j$ varies over the taxa; the basis functions themselves do not evolve in our model.

In section \ref{sec:PhyloGP} we address the problem of estimating the statistical structure of the mixing coefficients by performing phylogenetic Gaussian process regression (PGPR) on each of the rows of $\hat{X}$ separately. This corresponds to assuming independence between the rows (i.e. that the coefficients of the different basis functions evolve independently). It is commonly argued in the quantitative genetics literature \cite{Butler04} that evolutionary processes can be modelled as 
Ornstein-Uhlenbeck (OU) processes.  Under these assumptions the estimation of the forward operator reduces to the estimation of a small vector $\gamma$ of parameters \cite{JM12}. In section \ref{sec:Simulation} we clarify the interpretation of these parameters in evolutionary contexts. The explicit PGPR posterior likelihood function is then used to obtain maximum 
likelihood (MLE) estimates for $\gamma$. The estimation of $\gamma$ is known to be a challenging statistical problem \cite{Beaulieu12}. We suggest an approach based on the principal of \emph{bagging} \cite{Breiman96} in section \ref{sec:HypEst}.

Our final contribution (section \ref{sec:AncRec}) addresses the problem of estimating the function-valued traits of ancestral taxa. The PGPR step above also returns a posterior distribution for the mixing coefficient of each basis function at each ancestral taxon in the phylogeny. At any particular ancestor the estimated basis functions may be combined statistically, using the posterior distributions of their respective mixing coefficients, to provide a function-valued posterior distribution. Since the univariate posterior distributions are Gaussian, and the mixing is linear, the posterior for the function-valued trait has a closed form representation as a Gaussian process (Eq. \ref{ancestorGP}) which provides a major analytical and computational advantage for the approach.
We can verify the methods proposed by using a PGP as a stochastic generative model. This simulates correlated function-valued traits across the taxa of $\bf{T}$. 
Given only the phylogeny and the function-valued traits of taxa at its tips, our estimates for $\hat{\phi}$ and the ancestral functions are then compared to the simulation.

\begin{figure}[!ht]
\includegraphics[width=.5\textwidth]{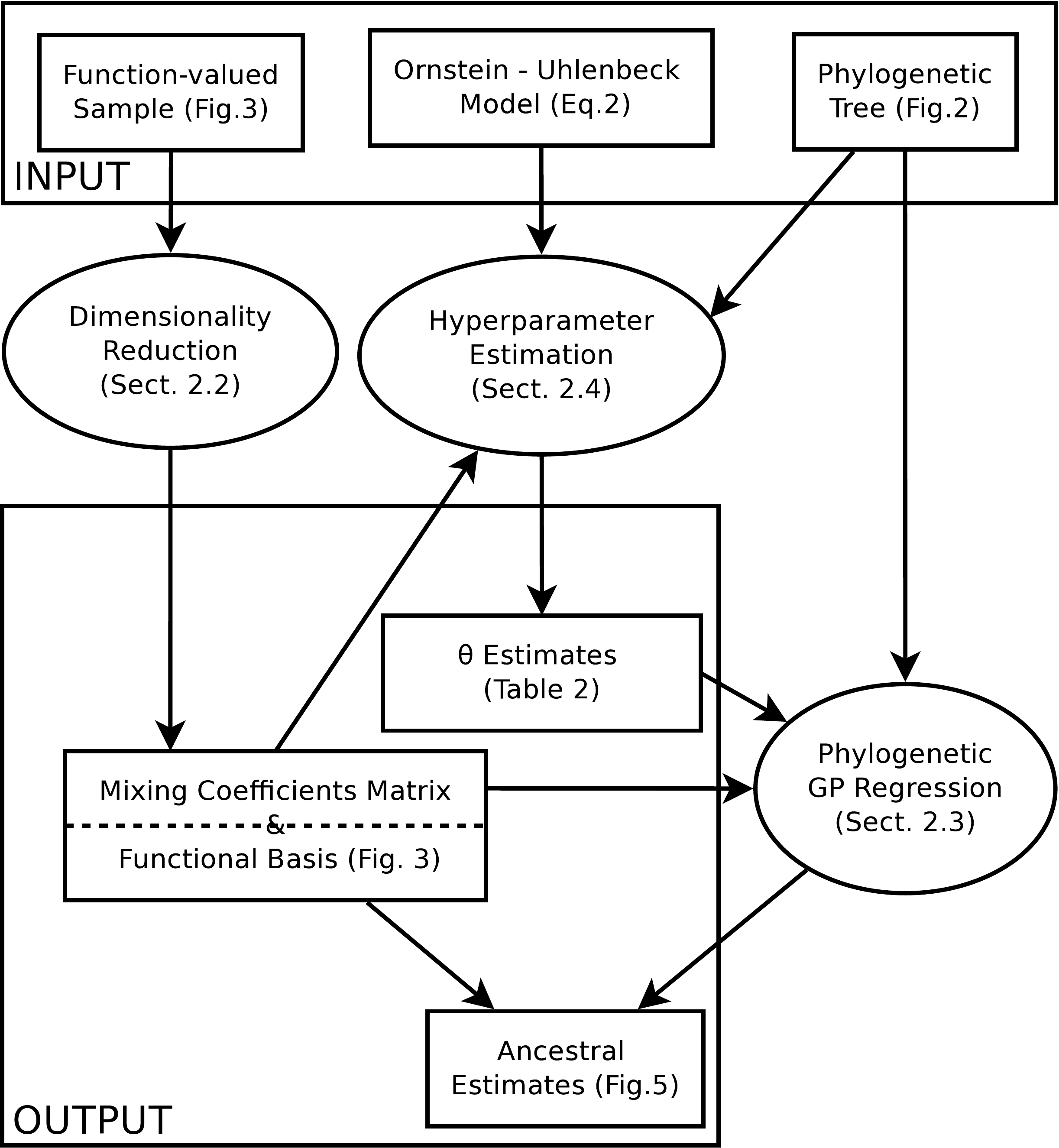}
\caption{The three methods presented in this paper (ovals) and their interrelationships.}
\label{FIG0}
\end{figure}

Overall, our three methods (in \ref{sec:DimensionRed}, \ref{sec:HypEst}, \ref{sec:AncRec}) appropriately combine developments in functional data analysis with the evolutionary dynamics of quantitative phenotypic traits, allowing nonparametric Bayesian inference from phylogenetically correlated function-valued traits. An outline of the framework presented in the current work can be found in Fig. \ref{FIG0}.

\section{Methods \& Implementation}
\label{sec:Methods}

\subsection{Artificial evolution of function-valued traits}
\label{sec:Simulation}

\begin{figure}[!t]
\includegraphics[width=.5\textwidth]{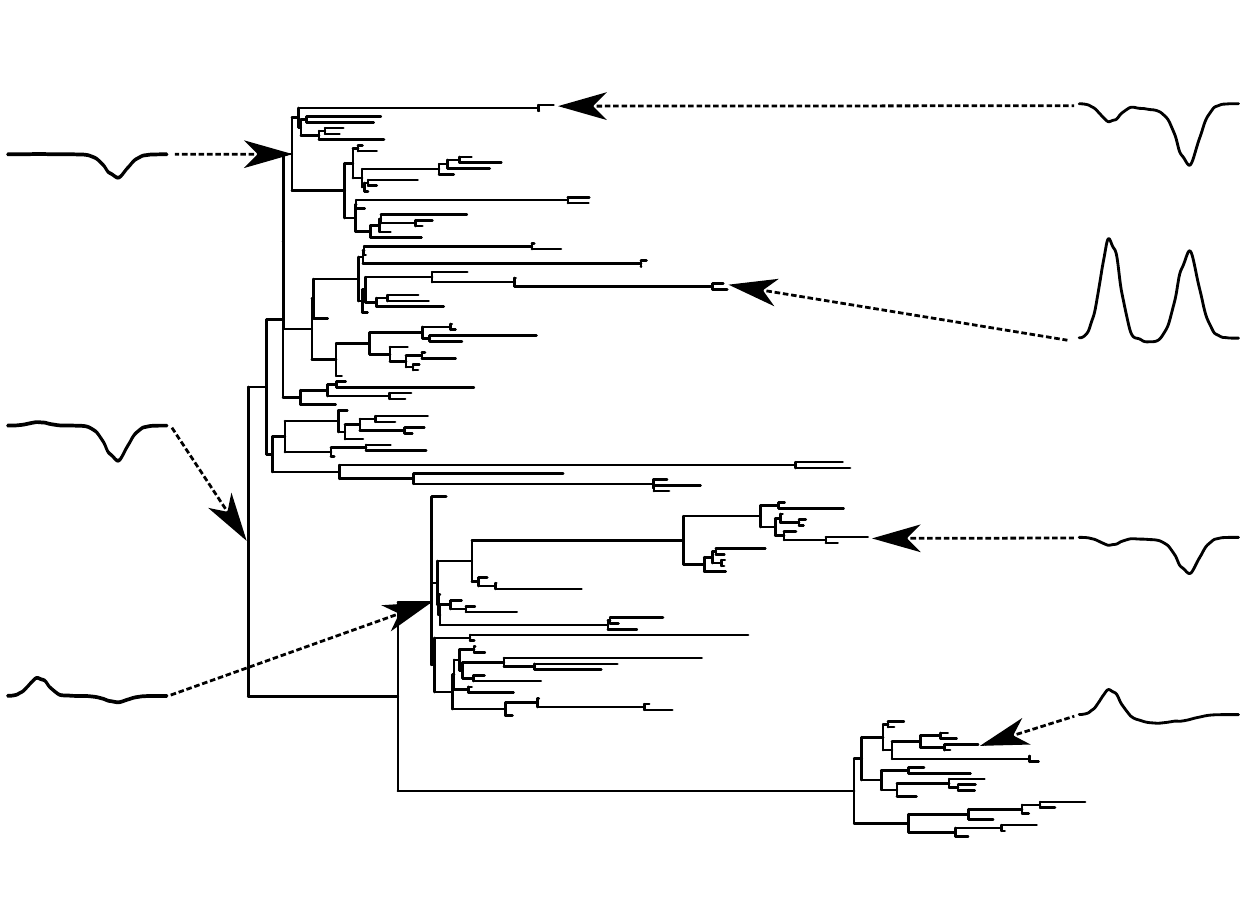} 
\caption{The random phylogenetic tree used \& examples of the function-valued traits shown at the tips (extant taxa) and the internal nodes (ancestral taxa). A subset of these is used in Fig. \ref{AncRecFig}.} \label{FIG_tree}
\end{figure}

We begin by generating a random phylogenetic tree $\bf{T}$ with 128 tips, shown in Fig. \ref{FIG_tree}. This fixes the experimental design for our simulation and inference, but further simulations given in the Supplementary Material confirm that the statistical performance of our methods is consistent across a range of choices for $bf{T}$. Branch length distributions are surprisingly consistent across organisms \cite{Venditti10}; branch lengths were drawn from the empirical branch length distribution\footnote{See Supplementary Material section 1} extracted from TreeFam 8.0 \cite{TreeFam}.

\begin{figure}[ht]
\includegraphics[width=.5\textwidth]{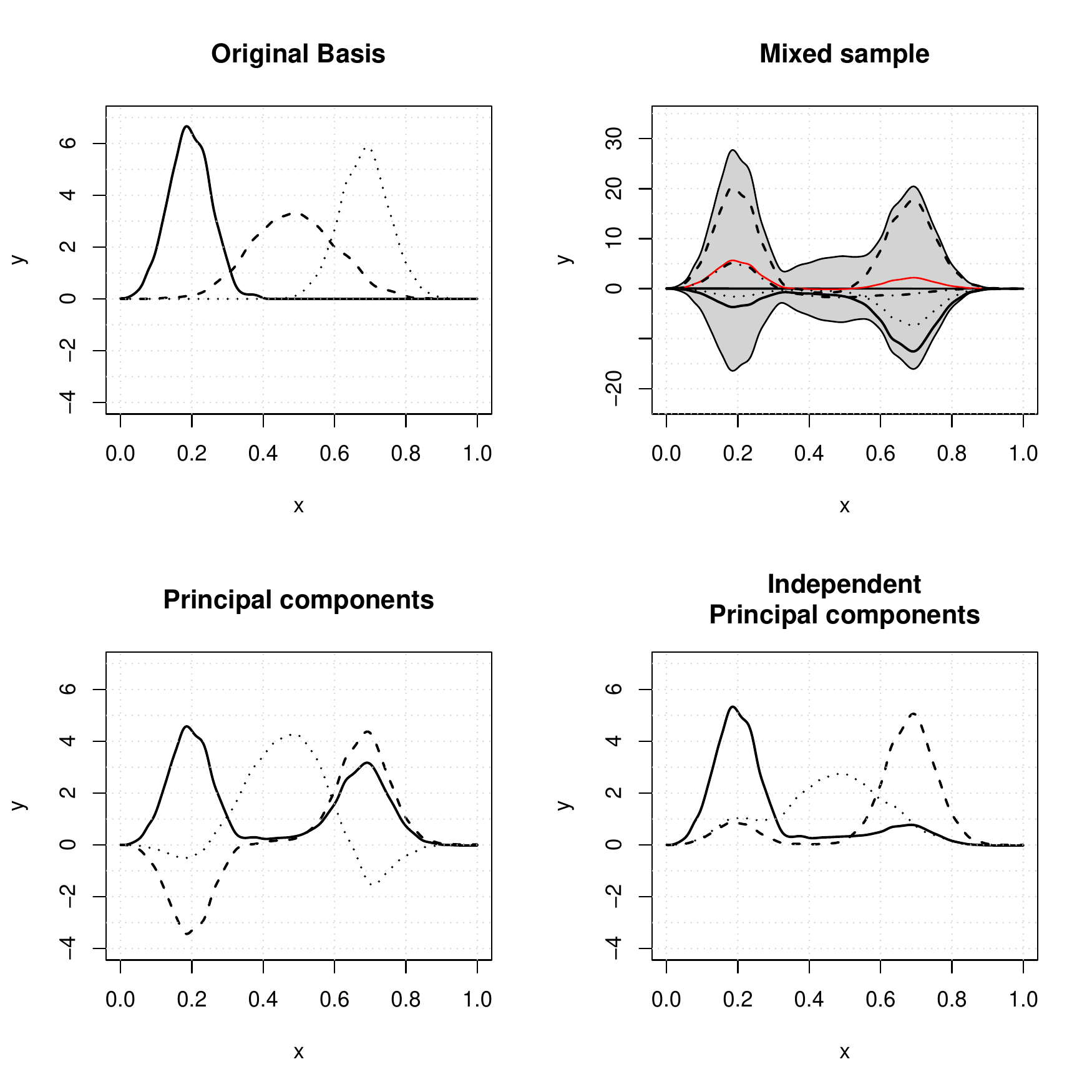}
\caption{From Top Left: Original Basis signals, $\phi$; Mixed Sample at the tips, $d$ (four individual function-valued traits are shown; red line and grey band show respectively the mean and two standard deviations for all 128 function-valued data at the tips); IPCA Basis, $\hat{\phi}$; PCA Basis.} \label{FIG_IPCA}
\end{figure}

Secondly we chose a basis $\phi$ in Eq. \ref{lip}. We have no reason {\em a priori} to suppose that this basis is orthogonal and, in general, there is no reason for our inference procedure to be sensitive to the particular shape of the basis functions. The three simple non-orthogonal, unimodal functions shown in Fig. \ref{FIG_IPCA} were therefore chosen as examples. For computational purposes each basis function was stored numerically as a vector of length 1024, so that the basis matrix $\phi$ was of size $3 \times 1024$ and its $i$th row stored the $i$th basis function.

Thirdly, different mixing coefficients were generated by a phylogenetic OU process for each basis function and stored in the respective row of $X$. Our modelling assumption is that the mixing coefficients for distinct basis functions $\phi_1,\phi_2,\phi_3$ are statistically independent of each other: in Eq. \ref{lip} this means that the rows of $X$ are independent. It is therefore sufficient to describe the stochastic process generating $X_i$, the $i$th row of $X$ with $i \in \{1,2,3\}$. We  calculated the mixing matrix at the 128 tip taxa so $X$ is of size $3\times 128$. The ``true'' ancestral values were established by generating phylogenetic Ornstein-Uhlenbeck (OU) processes over the whole phylogeny.  
The values of this process at tip taxa were stored in a row vector $\overline{X}_i$ ($\overline{X}_i$ is a simulation of the tip taxa mixing coefficients $X_i$ excluding the non-phylogenetic variation)
and its values at internal taxa were stored in a row vector $W_i$ for performance analyses in section \ref{sec:AncRec}. To simulate the additional effect of non-phylogenetic variation (due, for example, to measurement error or environmental effects), independent (that is, non-phylogenetic) variation was added to each entry of $\overline{X}_i$:
\[X_i = \overline{X}_i + {\bf \epsilon}_i\]
where $\epsilon_i$ is a $1 \times 128$ vector of independent Gaussian errors with mean 0 and variance $\sigma_n^i$ and finally the matrix multiplication in Eq. \ref{lip} was performed to obtain the simulated data $d$. The `extant' function-valued trait at tip taxon $j$ is thus  $\sum_{i=1}^3 X_{ij}\phi_i$ (a vector of length 1024), while the ancestral function-valued trait at internal taxon $g$ is $\sum_{i=1}^3 W_{ig}\phi_i$. The ancestral function-valued traits therefore exhibit only the phylogenetic part of simulated variation, while the extant function-valued traits exhibit both phylogenetic and non-phylogenetic variation. Of course, it is not possible to reconstruct non-phylogenetic variation using phylogenetic methods: we simulate non-phylogenetic variation only to demonstrate that it does not prevent the reconstruction of the phylogenetic part of variation for ancestral taxa in sections \ref{sec:DimensionRed} to \ref{sec:AncRec}.

We now comment on the specific parameters chosen for the phylogenetic OU processes above. As in  \cite{Hansen97} we refer to the {\em strength of selection parameter} $\alpha$ and the {\em  random genetic drift} $\sigma$: we add superscripts to these parameters to distinguish between the three different OU processes. With this notation, the mixing coefficients for the row $X_i$ have the following covariance function :
\begin{align}
 K_{\bf T}^i(\bt_1,\bt_2)=&  E[X_{ij} X_{ig}] \label{oucov}
\\  \quad =& (\sigma_f^{i})^2  \exp\left( -2\alpha^i D_T(\bt_j,\bt_g)\right) + (\sigma_n^i)^{2} \delta_{\bt_j,\bt_g}^e  \nonumber \end{align}
where $\sigma^i_f=\sqrt{\frac{(\sigma^i)^2}{2\alpha^i}}$ 
, $D_T(\bt_j,\bt_g)$ denotes the phylogenetic or patristic distance (that is, the distance in ${\bf T}$) between the $j$th and $g$th tip taxa, $\sigma_n$ is defined as above, and 
\begin{eqnarray*}
\delta_{t_j,t_g}^e = \left\{ \begin{array}{ll}1 & \text{iff } t_j = t_g \text{ and } t_j \text{ is a tip taxon}, \\
0 & \text{ otherwise}
\end{array}\right.
\end{eqnarray*}
adds non-phylogenetic variation to extant taxa as discussed above, ie. $\delta^e$ evaluates to 1 only for extant taxa, thus $\sigma_n$ quantifies within-species genetic or environmental effects and measurement error in the ith mixing coefficient.
We see from Eq. \ref{oucov} that the proportion of variation in the row $X_i$ attributable to the phylogeny is $\frac{(\sigma_f^i)^2}{(\sigma_f^i)^2 + (\sigma_n^i)^2}$. 


In the Gaussian process regression literature in Machine Learning, $\frac 1 {2\alpha}$ is equivalent to $\ell$, the characteristic length-scale \cite{Rasmussen06} of decay in the correlation function and in the following we work with the latter. 
For all of the OU processes we used characteristic length scales relative to 8.22, the maximum patristic distance ($\ell_{max}$) between two extant taxa for our simulated tree (Fig. \ref{FIG_tree}).
The values we used are given in Table \ref{Tab1}. In particular, $\sigma_f^i=0$ when $i=2$ and it follows that the characteristic length scale $\ell$ plays no role for this OU process, and equally we do not define the strength-of-selection parameter $\alpha^i$ when $i=2$.

\begin{table}[!ht]
\begin{center}
\begin{tabular}{cccc}
   $i$     &  $\sigma_{f}^i$ &  $\ell^i$  &  $\sigma_{n}^i$ \\ \hline
 1 & 2.5 & 6.17 & .5 \\
 2 & 0 & NA & 1 \\
 3 & 1.5 & 2.06 & .5 \\
\end{tabular}
\end{center}
\caption{The fixed values used for the parameters in Eq. \ref{oucov} to generate the mixing coefficients $X_{ij}$. Each row constitutes a value of $\gamma^i$. 6.17 \& 2.06 correspond to .75 and .25 of the tree's $\ell_{max}$ respectively. When $i=2$, $\ell^{i}$ is not  applicable since there is no phylogenetic variation in the sample. 
}
\label{Tab1}
\end{table}

\subsection{Dimensionality reduction and source separation for function-valued traits}
\label{sec:DimensionRed}

Given a dataset $d$ of function-valued traits, we would like to find appropriate estimates $\hat{X}$ and $\hat{\phi}$ of the mixing matrix $X$ and the basis set $\phi$ respectively. The first task is to identify a good linear subspace $S$ of the space of all continuous functions by choosing basis functions appropriately. The purpose is to work, not with the function-valued data directly, but with their projections in $S$. We may say that the chosen subspace $S$ is good if the projected data approximate the original data well while the number of basis functions is not unnecessarily large, so that $S$ has the `effective' dimension of the data.

We then face a linear inverse problem: given the dataset $d$ of function-valued traits, the task is to generate estimates $\hat{X}$ and $\hat{\phi}$ (Eq. \ref{lip}). This task is also known as {\em source separation} \cite{Hyvarinen00}, which has a variety of implementations making different assumptions about the basis $\phi$ and mixing coefficients $X$. One widely used approach is PCA \cite{Bishop06}, 
 which returns orthogonal sets of basis functions to explain the greatest possible variation. PCA has been extended to take account of phylogenetic relationships \cite{Revell09PCA}, however, if a sample of functions is generated by mixing non-orthogonal basis functions, the principal components of the sample (whether or not they account for phylogeny) will not equal the basis curves, due to the assumption of orthogonality: see Fig. \ref{FIG_IPCA}. 
In independent components analysis (ICA), the alternative assumption is made that the rows $X_i$ of $X$ are  statistically independent. This assumption fits more naturally with our modelling assumptions, since we assume that the rows $X_i$ are mutually independent \cite{Hyvarinen00}. ICA has proved fruitful in other biological applications \cite{Scholz04} as has passing the results of PCA to ICA, which has been termed IPCA \cite{Yao12}.

PCA \textit{is} an appropriate tool for identifying the effective dimension of a high-dimensional dataset \cite{Minka00}. So, to achieve both dimension reduction and source separation, we first applied PCA to the dataset $d$ (the 128 function-valued traits at the tips of ${\bf T}$) to determine the appropriate number of basis functions. The principal components were then passed to the \emph{CubICA} implementation of ICA \cite{Blaschke04}. CubICA returned a new set of basis functions (Fig. \ref{FIG_IPCA}, lower-right panel) which were taken as the estimated basis $\hat{\phi}$.

\subsection{Phylogenetic Gaussian process regression}
\label{sec:PhyloGP}

ICA also returns the estimated mixing coefficients at tip taxa, $\hat{X}$. Our next step was to perform PGPR \cite{JM12} separately on each row $\hat{X}_i$, assuming knowledge of the phylogeny $\bf{T}$, in order to obtain posterior distributions for all mixing coefficients 
throughout the tree $\bf{T}$.

Gaussian process regression (GPR) \cite{Rasmussen06} is a flexible Bayesian technique in which prior distributions are placed on continuous functions. Its range of priors includes the Brownian motion and Ornstein-Uhlenbeck (OU) processes, which are by far the most commonly used models of character evolution \cite{Hansen96,Butler04}.
Its  implementation is particularly straightforward since the posterior distributions are also Gaussian processes and have closed forms. We now give a brief exposition of GPR, using notation standard in the Machine Learning literature (see, for example, \cite{Rasmussen06}).

A Gaussian process may be specified by its mean surface and its covariance function $K(\gamma)$, where $\gamma$ is a vector of parameters. Since the components of $\gamma$ parameterise the prior distribution, they are referred to as {\em hyper}parameters. The Gaussian process prior distribution is denoted
\[f\sim\mathcal{N}(0,K(\gamma))\]

If $x^*$ is a set of unobserved coordinates and $x$ is a set of observed coordinates, the posterior distribution of the vector $f(x^*)$ given the observations $f(x)$ is
\begin{align}\label{dangermouse}
f(x^*)|f(x)  \sim \mathcal N (A,B)
\end{align}
where
\begin{align}
A =& K(x^*,x,\gamma)K(x,x,\gamma)^{-1}f(x), \label{boselecta} \\
B =& K(x^*,x^*,\gamma)\notag\\ -& K(x^*,x,\gamma)K(x,x,\gamma)^{-1}K(x^*,x,\gamma)^{T} \label{chas and dave}
\end{align}
and $K(x^*,x,\gamma)$ denotes the $|x^*| \times |x|$ matrix of the covariance function $K$ evaluated at all pairs $x^*_i \in X^*, x_j \in X$. Equations \ref{boselecta} and \ref{chas and dave} convey that the posterior mean estimate will be a linear combination of the given data and that the posterior variance will be equal to the prior variance minus the amount that can be explained by the data. Additionally, the log-likelihood of the sample $f(x)$ is
\begin{align}\label{log1}
\log p(f(x)|\gamma) = &-\frac{1}{2}f(x)^T K(x,x,\gamma)^{-1}f(x) \notag\\
- &\frac{1}{2}\log(det(K(x,x,\gamma)))- \frac{|x|}{2} \log 2\pi.
\end{align}
It can be seen from Eq. \ref{log1} that the maximum likelihood estimate is subject both to the fit it delivers (the first term) and the model complexity (the second term). 
Thus, Gaussian process regression is non-parametric in the sense that no assumption is made about the structure of the model: the more data gathered, the longer the vector $f(x)$, and the more intricate the posterior model for $f(x^*)$.

PGPR extends the applicability of GPR to evolved function-valued traits.
A \emph{phylogenetic} Gaussian process is a Gaussian process indexed by a phylogeny ${\bf T}$, where the function-valued traits at each pair of taxa are conditionally independent given the function-valued traits of their  common ancestors. When the evolutionary process has the same covariance function along any branch of $\bf{T}$ beginning at its root (called the {\em marginal covariance function}), these assumptions are sufficient to uniquely specify the covariance function of the PGP, $K_{\bf T}$. As we assume that $\bf{T}$ is known in our inverse problem, the only remaining modelling choice is therefore the marginal covariance function. As can be seen from Eq. \ref{oucov}, $K$ is a function of patristic distances on the tree rather than  Euclidean distances as standard in spatial GPR. 

In comparative studies, where one has observations at the tips of ${\bf T}$, the covariance function $K_{\bf T}$ may be used to construct a Gaussian process prior for the function-valued traits, allowing functional regression. In the model that we use 
 this is equivalent to specifying a Gaussian prior distribution for the mixing coefficients $Y_{ij}$ and $X_{ij}$. This may be done by regarding the row vectors $Y_i$  and $X_i$ as observations of a univariate PGP. As noted in \cite{JM12}, if we assume that the evolutionary process is Markovian and stationary then the modelling choice vanishes and the marginal covariance function is specified uniquely: it is the stationary OU covariance function. If we also add explicit modelling of non-phylogenetically related variation at the tip taxa, the univariate prior covariance function has the unique functional form presented in Eq. \ref{oucov}. We do not assume knowledge of the parameters of Eq. \ref{oucov} however: their estimation is the subject of the next section.

\subsection{Hyperparameter estimation}
\label{sec:HypEst}
Since the posterior distributions returned by PGPR depend on the hyperparameter vector $\gamma$, we must estimate $\gamma$ in order to reconstruct ancestral function-valued traits, and the estimation procedure should correct for the dependence due to phylogeny. Maximum likelihood estimation (MLE) of the phylogenetic variation, non-phylogenetic variation and characteristic-length-scale hyperparameters $\sigma_f^i$, $\sigma_n^i$ and $\ell^i$ respectively may be attempted numerically using the explicit prior likelihood function (Eq. \ref{log1}). Because estimating $\sigma_f^i$ and $\ell^i$ alone is challenging \cite{Beaulieu12} (although the estimation improves significantly with increased sample size), and we have further increased the challenge by introducing non-phylogenetic variation, we propose an improved estimation procedure using the machine learning technique {\em bagging} \cite{Breiman96}, which a member of the \emph{boosting} framework \cite{Bishop06}. 
We show that these estimates may be further improved if one knows the value of the ratio $\frac{(\sigma_f)^2}{(\sigma_n)^2}$, which is closely related to Pagel's $\lambda$ \cite{Pagel97}.


Bagging (bootstrap aggregating) seeks to reduce the variance of an estimator by generating multiple estimates and averaging. It is simple to implement given an existing estimation procedure: one adds a loop front end that selects a bootstrap sample and sends it to the estimation procedure and a back end that aggregates the resulting estimates \cite{Breiman96}. We generated 100 (sub)trees of 100 taxa by sampling without replacement our original 128 taxa tree, obtained the MLE for $\gamma$ on each subtree, and  averaged these estimates to obtain the aggregated estimate $\hat{\gamma}$. Our results are shown in Table \ref{Tab2}: for $i=1$ and $i=3$, given our moderate sample size (128 taxa), the accuracy of these results is at least in line with the state of the art \cite{Beaulieu12} despite the additional challenge posed by non-phylogenetic variation. For $i=2$, where phylogenetic variation is absent from the generative model ($\sigma_f^i$=0), our estimation procedure indicates its absence by returning estimates for $\ell^i$ whose magnitude is unrealistically small for the examined tree (less than the 1st percentile of the tree's patristic distances). 
Commenting further on this matter, exceptionally \emph{small} characteristic length-scales relative to the tree patristic distances, as seen here, practically suggest taxa-specific phylogenetic variation, ie. non-phylogenetic variation. This holds also in its reverse: exceptionally \emph{large} characteristic length-scales suggest a stable, non-decaying variation across the examined taxa that is indifferent to their patristic distances, again suggesting the absence of phylogenetic variance among the nodes.

\begin{figure*}[!htb]
\includegraphics[width=.95\textwidth]{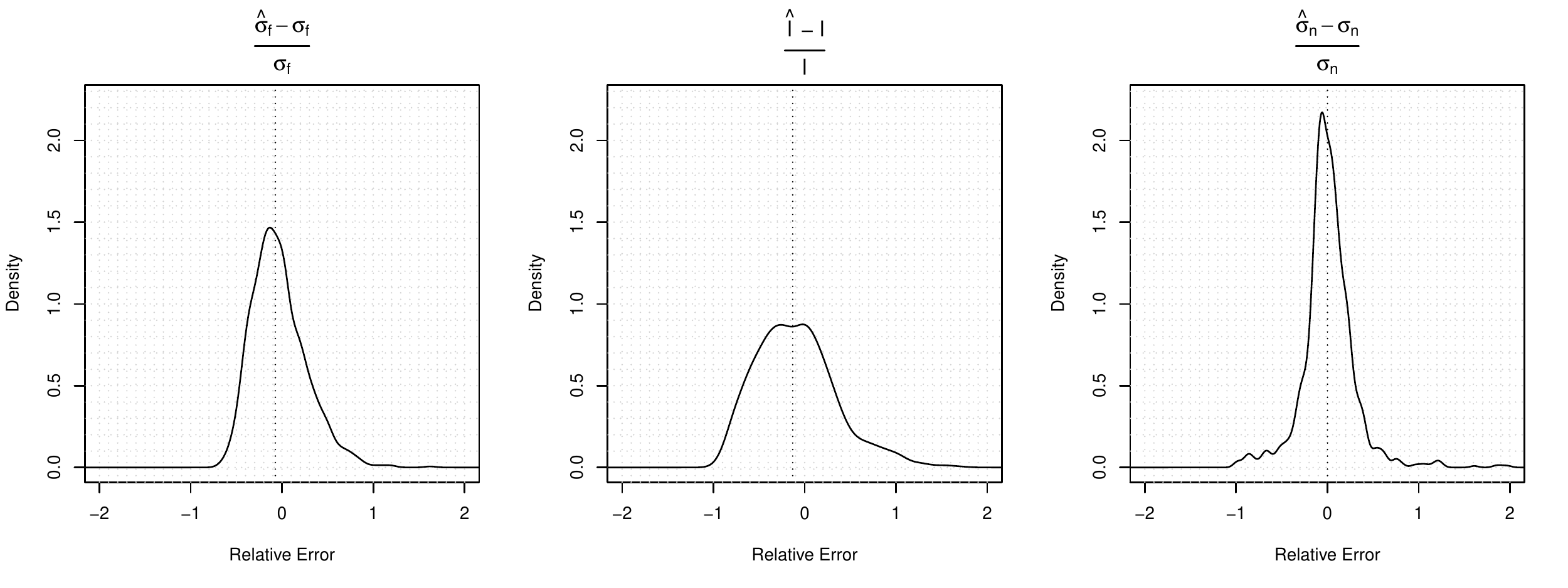}
\caption{Kernel Density estimates of the relative errors in 1024 runs of the $\gamma$ estimation procedure, each time for a different tree, a different set of mixing coefficients and a different set of parameters in $\gamma$; no components of $\gamma$ are assumed to be known beforehand. Estimation results are commented on in the Discussion. 
The median values shown by the dotted line are (-0.073, -0.131 and 0.001) respectively.
} \label{Fig_BaggingResults}
\end{figure*}

To assess the robustness of this hyperparameter estimation method we performed 1024 simulations, randomly regenerating the tree and parameter vector $\gamma$ each time\footnote{See Supplementary Material section 2}. The accuracy of these estimates is shown in Fig. \ref{Fig_BaggingResults}. Improved results when the ratio $\frac{(\sigma_f)^2}{(\sigma_n)^2}$ is known {\em a priori} (for example, through knowledge of Pagel's $\lambda$) are also given in the supplementary material (Sect. 2 \& 3). Our ultimate aim is ancestor reconstruction rather than hyperparameter estimation {\em per se}, and this is the subject of the next section.

\begin{table}[!ht]
\begin{center}
\begin{tabular}{cccc}
   $i$     &  $\hat{\sigma}_{f}^i$ &  $\hat{\ell}^i$  &  $\hat{\sigma}_{n}^i$ \\ \hline
 1 & 3.41 (.62) & 2.83 (.47)& 0.78  (.47) \\
 2 & 0.55 (.33)& 0.05 (.02) & 0.84  (.34) \\
 3 & 2.83 (.33) & 2.06 (.50) & 0.73 (.29) \\  
\end{tabular}
\end{center}
\caption{The bagging estimates for the hyperparameters in Eq. \ref{oucov} (standard deviations of bagging estimates in parentheses). Each row corresponds to a given estimate of the vector $\hat{\gamma}^i$. These estimates provide the maximum likelihood value for Eq. \ref{log1} and are comparable with the original ones from Table \ref{Tab1}.}\label{Tab2}
\end{table}


\subsection{Ancestor reconstruction}
\label{sec:AncRec}

Having generated function-valued data (Sect. \ref{sec:Simulation}), extracted mixing coefficients $\hat{X}$ (Sect. \ref{sec:DimensionRed}) and performed hyperparameter estimation (Sect. \ref{sec:HypEst}), we may now perform PGPR (Sect. \ref{sec:PhyloGP}) on each row $\hat{X}_i$, to obtain the univariate Gaussian posterior distribution for the mixing coefficient $W_{i\bt^*}$ at any internal taxon $\bt^*$. As discussed in Sect. \ref{sec:PhyloGP}, the Gaussian process prior distribution has covariance function (Eq. \ref{oucov}). We have assessed the accuracy of our bagging estimate $\hat{\gamma}$ in Sect. \ref{sec:HypEst} and we now 
substitute $\hat{\gamma}^i$ into Eq. \ref{oucov}. Taking a simple and direct approach, our estimate $\hat{\phi}$ obtained in Sect. \ref{sec:DimensionRed} may then be substituted into Eq. \ref{lip} to obtain the function-valued posterior distribution $f_{\bt^*}$ for the function-valued trait at taxon $\bt^*$. Since our estimated basis functions are stored numerically as vectors of length 1024, this gives the same discretision for the ancestral traits.

\begin{figure*}[!ht]
\includegraphics[width=.95\textwidth]{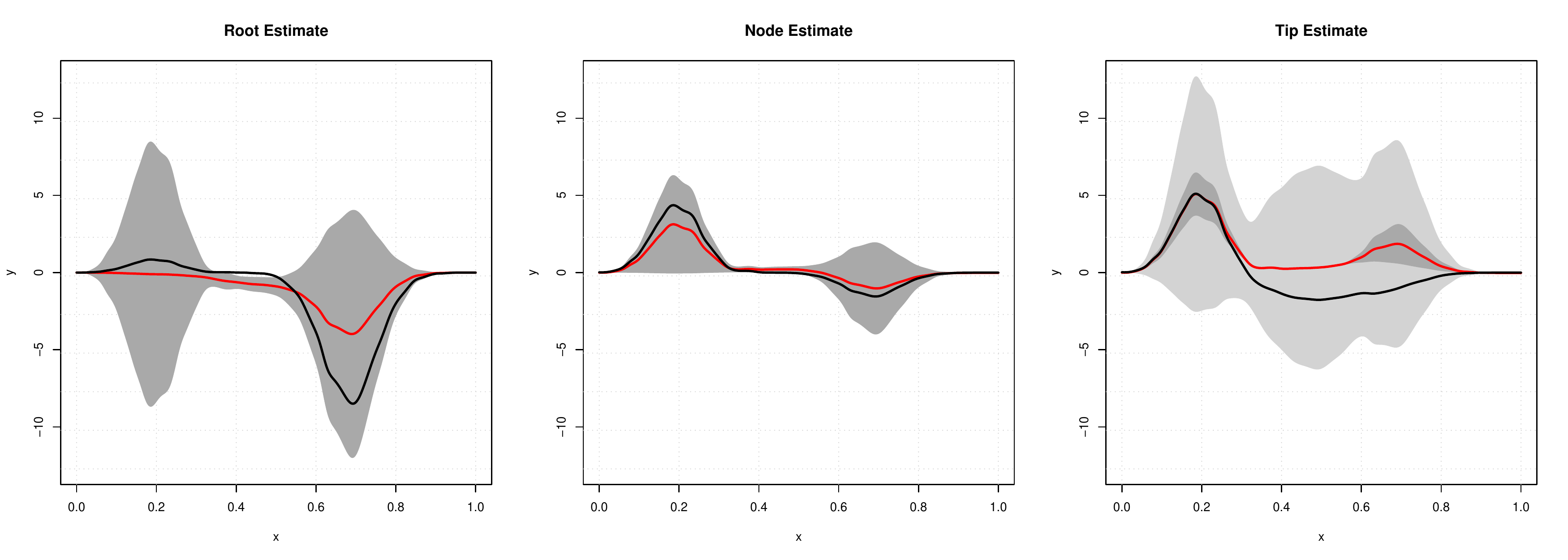}
\caption{Posterior distributions at three points in the phylogeny using the estimated $\hat{\phi}$ and $\hat{\gamma}$.
 The prediction made by the regression analysis is shown via the posterior mean (red line), the component of posterior variance due to phylogenetic variation (two standard deviations, dark grey band) and non-phylogenetic variation (two standard deviation, light grey band). The black line shows the simulated data enabling visual validation of the ancestral predictions. In the right panel, the black line is the training data at a tip taxon the red line and dark grey band represent the posterior distribution of its phylogenetic component, while the light grey band represents the estimated magnitude of non-phylogenetic variation. The root and internal taxon here are the same as those indicated in Fig.\ref{FIG_tree} \& \ref{FIG_IPCA}, and the tip is the second from bottom on the same figure.} \label{AncRecFig}
\end{figure*}

Conditioning 
 on our estimated mixing coefficients $\hat{X}_i$ for the tip taxa, the posterior distribution of $W_{i \bt^*}$ is \[W_{i \bt^{*}}\sim\mathcal{N}(\hat{A_i},\hat{B_i})\]
where the vector $\hat{A_i}$ and matrix $\hat{B_i}$ are obtained from Eq.'s \ref{boselecta} and \ref{chas and dave}, taking $x=\hat{X}_i$, $x^*=W_{i\bt^*}$ and $\hat{\gamma^i}$ respectively for our observation coordinates, estimation coordinates and hyperparameter vector. Since our prior assumption is that the rows of $X$ are statistically independent of each other, it follows from Eq. \ref{lip} that
\begin{equation}\label{ancestorGP}
f_{\bt^*}\sim\mathcal{N}(\Sigma_{i=1}^k \hat{A_i} \hat{\phi_i},  \Sigma_{i=1}^k \hat{\phi}_i^T \hat{B_i} \hat{\phi_i})
\end{equation}
The marginal distributions of this representation (mean and standard deviation) are shown in Fig. \ref{AncRecFig}.


Fig. \ref{AncRecFig} compares the function-valued estimates $\hat{f_{\bt^*}}$ to the simulated function-valued traits at the root (left panel), an internal node (centre), and at a tip (right panel). In the centre and left panels the simulated function-valued data is shown in black, and can be seen typically to lie within two posterior standard deviations. In the right panel, the black line is the observed function-valued trait at that tip: the red line and dark grey band represent the posterior distribution of its phylogenetic component, and the light grey band represents the estimated magnitude of the additional non-phylogenetic variation.
Uncertainty over the phylogenetic part of variation (dark grey band) decreases from root to tip, as all observations are at the extant tip taxa. We note that the posterior distributions, even at the root, put clear statistical constraints on the phylogenetic part of ancestral function-valued data: in this (admittedly simulated and highly controlled) setting we can reason effectively about ancestral function-valued traits.

\section{Discussion}

In Sec. \ref{sec:Simulation} we have appealed to Eq. \ref{lip} in the setting of mathematical inverse problems where, given data $d$, the challenge is to infer a forward operator $G$ and model $\phi$ such that:
\begin{equation} d = G(\phi)
\end{equation}
and such problems are typically under-determined and require additional modelling assumptions \cite{Jaynes84}. 
Given a phylogeny $\bf{T}$ and function-valued data $d$ at its tips, we wish to infer the forward operator $G_{\bf{T}}$ and model $\phi$ such that
\begin{equation} d = G_{\bf{T}}(\phi)
\end{equation}

When the data $d$ are a small number of correlated factors per tip taxon, a variety of statistical approaches are available (e.g. see \cite{Salamin10},  \cite{Hadfield10}). When the data are functions, the phylogenetic Gaussian processes (PGP) \cite{JM12,Kerr12} has been proposed as the forward operator and this is the approach we have taken in this work.

Our dimensionality reduction methodology in Sect. \ref{sec:DimensionRed} can be easily varied or extended. For example, any suitable implementation of PCA may be used to perform the initial dimension reduction step: in particular, if the data have an irregular design (as happens frequently with function-valued data), the method of Yao et al. \cite{Yao05} may be applied to account for this; the ICA step then proceeds unchanged. We also note that while we find the {\em CubICA} implementation of ICA to be the most successful in our signal separation task, other implementations like {\em FastICA} \cite{Hyvarinen00} or {\em JADE} \cite{Cardoso99} can also be employed.
In general, ICA gives rows $\hat{X}_i$ of the estimated mixing matrix that are maximally independent under a particular measure of independence involving, for example, higher sample moments or mutual information, 
in order to approximate the solution of the inverse problem in Eq. \ref{lip} under our assumption of independence between the rows of $X$. 
PCA and ICA have different purposes (respectively, orthogonal decomposition of variation and separation of independently mixed signals) and we employ them sequentially in IPCA. IPCA is nonparametric and, in particular, both distributionally and phylogenetically agnostic. This means that unlike PCA, IPCA is robust to non-Gaussianity in the data and, unlike phylogenetically corrected PCA, IPCA is robust to mis-specification of the phylogeny and to mixed phylogenetic and non-phylogenetic variation in the data: any of these can be features of biological data.

It can be seen in Fig. \ref{Fig_BaggingResults} that the estimation of $\ell$ is more challenging than the estimation of $\sigma_n$ or $\sigma_f$, having greater bias and variance. This corresponds to the documented difficulty of estimating the parameter $\alpha$ in the Ornstein-Uhlenbeck model, particularly for smaller sample sizes.
Our work on  hyperparameter estimation in Sec. \ref{sec:HypEst} mitigates these difficulties due to small sample size \cite{Beaulieu12,Collar09} by employing bagging in order to bootstrap our sample. Somewhat unintuively, bagging ``works'' exactly because the subsample $\hat{\gamma}$ estimates are variable and thus we avoid overfitted final estimates\footnote{See Supplementary Material Section 2}. 
Conceptually our work on hyperparameter estimation, when taken together with Sec. \ref{sec:DimensionRed}, relates to the character process models of \cite{Pletcher1999} and orthogonal polynomial methods of \cite{Kirkpatrick89}, which give estimates for the autocovariance of function-valued traits. Writing out Eq. \ref{lip} for a single function-valued trait (at the $j$th tip taxon, say), our model may be viewed as
\begin{equation}
f(x) = \sum_{i=1}^3 g_{ij} \phi_i(x) + \sum_{i=1}^3 e_{ij} \phi_i(x)
\end{equation}
where the mixing coefficient $X_{ij}$ has been expressed as the sum of $g_{ij}$, the genetic (i.e. phylogenetic) part of variation, plus $e_{ij}$, the non-phylogenetic (eg. environmental) part of variation, just as in these references. 
Then the autocorrelation of the function-valued trait is
\begin{equation}\label{buzz}
E[f(x_1)f(x_2)] = \sum_{i=1}^3 \left((\sigma^f_{i})^2 + (\sigma^n_i)^2\right) \phi_i(x_1)\phi_i(x_2)
\end{equation}
The estimates of $\sigma^f_{i}$ and $\sigma^n_{i}$ obtained in section \ref{sec:HypEst} may be substituted into Eq. \ref{buzz} to obtain an estimate of the autocovariance of the function-valued traits under study. This estimate has the attractions both of being positive definite (by construction) and of taking phylogeny into account.

Various frameworks exist which could be used to generalise the method presented in Sec. \ref{sec:HypEst}, to model heterogeneity of evolutionary rates along the branches of a phylogeny \cite{Revell09} or for multiple fixed \cite{Butler04} or randomly  evolving \cite{Hansen08, Beaulieu12} local optima of the mixing coefficients. For the stationary OU process the optimum trait value appears only in the mean, and not in the covariance function, and so does not play a role as a parameter in GPR (see \cite{Rasmussen06}). We have not implemented such extensions here, effectively assuming that a single fixed optimum is adequate for each mixing coefficient. Nonetheless our framework is readily extensible to include such effects, either implicitly through branch-length transformations \cite{Pagel99}, or explicitly by replacing the OU model with the more general Hansen model \cite{Hansen08}.


R Code for the IPCA, ancestral reconstruction and hyperparameter estimation is available from \\ \url{https://github.com/fpgpr/}

\bibliography{Bibliography}
\bibliographystyle{vancouver}

\end{document}